\magnification = \magstep1
\overfullrule=0pt
\raggedbottom
\baselineskip = 3ex
\raggedbottom
\font\eightpoint=cmr8

\font\fivepoint=cmr5
\headline={\hfill{\fivepoint EHL\ 3 Dec, 1996}}

\def\uprho{{\raise 1pt\hbox{$\rho$}}}
\def\C{{\bf C}}

\def\B{{\bf B}}
\def\D{{\cal D}}
\def\R{{\bf R}}

\def\T{{\cal T}}

\def\pslash{{\bf p\hskip-.16cm {\bf /}}}
\def\Aslash{{\bf A\hskip-.2cm/}}
\def\Oslash{{O\hskip-.22cm/}}
\def\A{{\bf A}}
\def\p{{\bf p}}

\newcount\chno \chno=0
\newcount\equno  
\newcount\refno \refno=0

\font\chhdsize=cmbx12 at 14.4pt

\def\startbib{\def\biblio{\bigskip\medskip
  \noindent{\chhdsize References.}\bgroup\parindent=2em}}
\def\endbib{\edef\biblio{\biblio\egroup}}
\def\reflbl#1#2{\global\advance\refno by 1
  \edef#1{\number\refno}
    \global\edef\biblio{\biblio\medskip\item{[\number\refno]}#2\par}}

\def\eqlbl#1{\global\advance\equno by 1
  \global\edef#1{{\number\chno.\number\equno}}
  (\number\chno.\number\equno)}

\centerline{\chhdsize STABILITY OF MATTER IN MAGNETIC FIELDS}
\bigskip\bigskip \bigskip

\centerline{Elliott H. Lieb$^*$} \vfootnote{}
{\eightpoint 
* Work partially
supported by U.S. National Science Foundation grant PHY 95-13072.}
\centerline{Departments of Physics and Mathematics }
\centerline{Princeton University}
\centerline{P.O. Box 708, Princeton, NJ  08544-0708, USA}

\noindent
{\baselineskip=2.5ex 
\vfootnote{}{ \eightpoint  \copyright 1996 by the author.
Reproduction of this article, in its entirety, by any means is permitted
for non-commercial purposes.}}

{\baselineskip=2.5ex
\vfootnote{}{ \eightpoint Lecture given at the Euroconference on
Correlations in Unconventional Quantum Liquids, Evora, Portugal,
October, 1996}}
\bigskip\bigskip

\bigskip\noindent

{\bf ABSTRACT:} The proof of the stability of matter is three decades
old, but the question of stability when arbitrarily large magnetic
fields are taken into account was settled only recently. Even more recent
is the solution to the question of the stability of relativistic matter
when the electron motion is governed by the Dirac operator (together
with Dirac's prescription of filling the ``negative energy sea"). When
magnetic fields are included the question arises whether it is
better to fill the negative energy sea of the free Dirac operator or of
the Dirac operator with magnetic field. The answer is found to be that
the former prescription is unstable while the latter is stable.  

\bigskip\bigskip \bigskip

\startbib
\reflbl\LSS{E.H. Lieb, H. Siedentop and J.P. Solovej,  {\it Stability 
and Instability of Relativistic Electrons in Classical Electromagnetic
fields}, Jour. Stat. Phys., (in press).}

\reflbl\DL{F.J. Dyson and A. Lenard, J. Math. Phys. {\bf 8}, 423
(1967); {\bf 9}, 698 (1967). }

\reflbl\LT{E.H. Lieb and W.E. Thirring, Phys. Rev. Lett. {\bf 35}, 687
(1975); Errata {\bf 35}, 1116 (1975).}

\reflbl\LR{E.H. Lieb, Rev. Mod. Phys. {\bf 48}, 553 (1976). }

\reflbl\LG{E.H. Lieb, Bull. Amer. Math. Soc. {\bf 22}, 1 (1990).}

\reflbl\AHS{J. Avron, I. Herbst and B. Simon, Commun. Math. Phys.
{\bf 79}, 529 (1981).}

\reflbl\FLL{J. Fr\"ohlich, E.H. Lieb and M. Loss, Commun. Math. Phys. 
{\bf 104}, 251 (1986);
E.H. Lieb and M. Loss, Commun. Math. Phys. {\bf 104}, 271 (1986).}

\reflbl\LY{M. Loss and H-T. Yau, Commun. Math. Phys. {\bf 104}, 
283 (1986).}

\reflbl\LLSA{E.H. Lieb, M. Loss and J.P. Solovej, Phys. Rev. Lett., 
{\bf 75}, 985 (1995).}

\reflbl\K{T. Kato, {\it Perturbation Theory for Linear Operators} in
Grundl. der mathem. Wissen., {\bf 132}, Springer (1966).}

\reflbl\H{I. Herbst, Commun. Math. Phys. {\bf 53}, 285 (1977).}

\reflbl\C{J.G. Conlon, Commun. Math. Phys. {\bf 94}, 439 (1984).}

\reflbl\LiY{E.H. Lieb and H-T. Yau, Commun. Math. Phys. {\bf 118}, 
177 (1988).}

\reflbl\LLSB{E.H. Lieb, M. Loss and H. Siedentop, {\it Stability of
Relativistic Matter via Thomas-Fermi Theory}, Helv. Phys. Acta,
{\bf 69}, No. 5/6, 1996 (in press).} 

\reflbl\EPS{D. Evans, P. Perry and H. Siedentop, 
Commun. Math. Phys. {\bf 178}, 733 (1996).}

\endbib

\chno=1
\bigskip\noindent
{\chhdsize I. Introduction}
\bigskip

This will be a brief summary of recent work which solves the problem of
the stability of matter in magnetic fields. The first part, which was
the one reported in the Evora conference, concerns the usual
Schr\"odinger equation with nonrelativistic kinetic energy, but with a
Zeeman term $\sigma \cdot \B$.  The second part is about some very
recent work [\LSS], not reported at the conference, on the relativistic
problem with kinetic energy given by the Dirac operator. In both cases
the electrons are treated quantum mechanically while the fields are
regarded as classical fields.  

It has been well understood for some time, starting with the work of
Dyson and Lenard in 1967 [\DL], that matter consisting of $N$ electrons
and $K$ static (but arbitrarily positioned) nuclei, and governed by
nonrelativistic quantum mechanics, is stable. (See also [\LT] and see
[\LR], [\LG] for reviews.)  This means that the ground state energy (more
generally, the bottom of the spectrum, in case there is no lowest
eigenvalue) is finite and is bounded below by a universal constant times
the number of particles, i.e., $E_0 \geq const. (N+K)$. The reason the
nuclei are taken to be static is that they are so massive compared to
the electron that if the nuclear mass (or the nuclear radius of
$10^{-13}$cm, for that matter) played any significant role  then
ordinary matter would have to look very different from what it does.  

The Hamiltonian of our system in suitable units is 
$$
H_{\rm nonrelatvistic}= \sum_{i=1}^N p_i^2 +\alpha V_{\rm c} \ ,
\eqno\eqlbl\aa
$$
where $\alpha=e^2/\hbar c$ is the fine-structure constant and
$\p=-i\nabla$, as usual.
$V_{\rm c}$ is the Coulomb potential of $K$ fixed nuclei with
nuclear charge $Ze$ (we take all the charges to be the same for
simplicity only), with locations  $R_j$ in $\R^3$, and with $N$
electrons.  

$$
V_{\rm c} = - Z\sum_{i=1}^N\sum_{j=1}^K |x_i-R_j|^{-1} +
\sum_{1 \le i<j \le N} |x_i-x_j|^{-1}
+ Z^2 \sum_{1 \le i<j \le K}|R_i-R_j|^{-1}. \eqno\eqlbl\ab
$$

We should like to generalize this stability question in two directions.
One direction is to add magnetic fields and the other is to make the
kinetic energy relativistic, which means replacing $p^2$ by $\sqrt{p^2
+m^2} -m$ or, equivalently and more simply by $|\p|$ (since the
difference between these two expressions is bounded by $m$, and hence
the change in energy is at most a term proportional to the number of
electrons). It will turn out that these two problems (magnetic fields
and relativity) are not as unrelated mathematically as one might think.
Later on, we shall generalize still further by replacing the operator
$|\p|$ by the Dirac operator; this is the content of the very recent
work mentioned above.  
\bigskip\bigskip

\noindent \chno=2 \equno=0
{\chhdsize II. Stability of non-relativistic matter with magnetic fields}
\bigskip

Let us begin with the magnetic field problem, and with a simple example
to illustrate its essence. We shall consider the nonrelativistic case
first.  We shall denote by $\A(x)$ the vector potential of a magnetic
field and $\B(x) = {\rm curl} \, \A(x)$ is the magnetic field itself.
Conceptually, we do not ask where this field comes from; it can be
external or it can be generated by the current of the electrons
themselves.  The action of the field on the orbital motion of the
electrons is given by the minimal substitution (recalling that the
electron's charge is negative)
$$   
\p \longrightarrow \p+\A(x). \eqno\eqlbl\ac
$$
This has little effect on the ground state energy of our Hamiltonian
in (\aa). Indeed, for one electron the field can only raise the energy.
For several electrons it can lower the energy but by a controllable
amount. In fact, all known proofs of the stability of matter hold with
constants unchanged by  the substitution (\ac).  

The problem comes in when we include the electron spin--field
interaction. Thus, we replace $(\p+\A)^2$ by the {\it Pauli operator}
(in which $\sigma$ denotes the triplet of Pauli matrices)
$$
\T_\A \equiv [\sigma \cdot (\p + \A)]^2= (\p+\A)^2+ \sigma \cdot \B .
\eqno\eqlbl\ad
$$

It is a fact, which is not entirely easy to prove [\AHS], 
that $E_0$ for a
hydrogen-like atom $H=\T_\A-Z /r$, behaves as $-(\ln B)^2$ for a large,
spatially uniform field $\B$. Thus, if we minimize $E_0$ over all
possible fields (since nature will presumably find that value of the
field that gives the lowest $E_0$) the minimum energy will be $-\infty$.
However, we forgot to include the self energy of the magnetic field
$$
H_{\rm field} \equiv [8\pi \alpha ]^{-1}\int B(x)^2\, d^3x. 
\eqno\eqlbl\ae
$$
(The unconventional factor $\alpha^{-1}$ in (\ae) comes from the system
of units we have been using.) Does the addition of $H_{\rm field}$
stabilize the system, i.e., is $E_0$ now bounded below, independent of
$\B$?

The answer, perhaps not unsurprisingly in view of the appearance of
$\alpha$ in (\ae), is yes {\it if and only if $\alpha$ is small
enough}.  More precisely it is the combination $Z\alpha^2$ that has to
be small enough.  This was proved in [\FLL], where it was shown that
the ``hydrogenic" Hamiltonian
$$
H_{hydrogenic} = \T_\A -Z\alpha /r + H_{\rm field}   \eqno\eqlbl\ag
$$
is bounded below if $Z\alpha^2\leq 9\pi^2/8$. Oddly, it was harder to
prove that the atom is {\it unstable} if $Z\alpha^2$ is too large; this 
fact relied on a difficult calculation in [\LY] that the Pauli kinetic
energy operator $\T_\A$ can have a zero mode for certain special choices 
of the $\A$ field. That is, the equation
$$
\T_\A \Psi =0             \eqno\eqlbl\ah
$$
has a solution for some magnetic field $\B$, whose 
energy $H_{\rm field}$ is finite,
and some square integrable, electron wave function $\Psi$. The ordinary 
kinetic energy $p^2$ does not have this remarkable property, i.e., the 
property that in a suitable field the electron can ``stand still".
(This happens for the uniform magnetic field, to be sure, but that
field has infinite energy.)

Obviously, the next question  is whether the full many-electron,
many-nucleus problem is stable when $p^2$ is replaced by $\T_\A$ and
with the addition of $H_{\rm field}$. Thus, we consider
$$
H_{\rm Pauli} \equiv \sum_{i=1}^N \T_\A(i) +\alpha V_{\rm c} + 
H_{\rm field} \ , \eqno\eqlbl\af
$$
and ask whether it is bounded below (for all $\A,\, N, \, K$ and all
locations of the nuclei) by some universal constant times $N+K$. This
problem was open for about 9 years and was first solved by Fefferman
(unpublished) for sufficiently small $\alpha, \, Z$ (which does not
include hydrogen in its range). A simple proof of stability with good
constants was given shortly thereafter in [\LLSA]: A representative
sufficient condition for stability in (\af), obtained in [\LLSA], is 
$$
Z\alpha^2 \leq 0.041 \quad \hbox{\rm and} \quad \alpha \leq 0.06.
\eqno\eqlbl\ai
$$
Another sufficient condition is $Z\leq 1050$ when $\alpha = 1/137$.
(A more recent result in [\LSS] improves this to $Z\leq 2265$.)
It is comforting to have this very large margin of stability in
(\af) because it means that the magnetic field-electron interaction 
can be treated perturbatively with confidence when $Z\leq 92$. 

It is important to understand why there is a bound on $\alpha$ alone in
(\ai), independent of the magnitude of $Z$ (as long as $Z>0$).  The
reason is that once one admits the possibility that a single atom can
collapse if $Z\alpha^2$ is too large (or $Z\alpha$ is too large in the
relativistic theories studied below) then one has to ask what prevents a
large number of small nuclei from coming together to form  a
supercritical nucleus? The answer is the Coulomb repulsion among the
nuclei, but this quantity is effectively governed by the parameter
$\alpha^{-3}$ (or $\alpha^{-1}$ in the relativistic theory) because the
nuclear repulsion is proportional to $Z^2\alpha =
(Z\alpha^2)^2/\alpha^3$.  In other words if $Z\alpha^2$ is truly the
relevant atomic parameter then it should be regarded as fixed, which
means that $\alpha^{-3}$ is the relevant parameter governing the nuclear
repulsion. Anyway, whether or not the reader believes this heuristic
discussion, the fact remains that the bound on $\alpha$ in (\ai) is
essential.

Two proofs of stability are given in [\LLSA]. The simplest uses an
ingredient that, at first sight appears to have nothing to do with the
magnetic problem---{\it the stability of relativistic matter}. 
Let us turn to that next.  
\bigskip\bigskip

\noindent \chno=3 \equno=0
{\chhdsize III. Relativistic matter without magnetic fields and  its
application to the nonrelativistic magnetic problem}
\bigskip

The relativistic Hamiltonian looks like (\aa), with the
substitution (\ac), but with $p^2$ replaced by $|\p|$. It contains no
$H_{\rm field}$ term. Thus, 
$$
H_{\rm rel} \equiv \sum_{i=1}^N |\p+\A|_i +\alpha V_{\rm c} \ . 
\eqno\eqlbl\aj
$$

Admittedly, this Hamiltonian is not well motivated physically because
the operator $|\p|$ is not local (unlike the Dirac operator). Minimal
substitution does preserve gauge invariance, however. There is no
spin-field interaction, as in the Pauli or Dirac Hamiltonians, but let
us take one step at a time.  It will turn out that the understanding of
(\aj) is central to the whole story.

There are two important remarks to be made about (\aj). The first is that
everything scales like $1/{\rm length}$ (if we let the $\A$ field scale
as $1/{\rm length}$ as well). Because of this, there are exactly two
possibilities for $E_0$ after we  minimize over all possibilities for
$\A$ and the nuclear locations:
$$
E_0 = 0   \quad\quad \hbox{\rm or }\quad \quad      E_0 = -\infty \ .       
      \eqno\eqlbl\ak
$$
(If we used $\sqrt{p^2+m^2} -m$ instead, then the first alternative, 
$E_0=0$  would be replaced by some negative number depending on
$m$. The condition on $\alpha $ for the first alternative to hold
does not depend on $m$.)

The second remark is that for the one-electron hydrogenic problem $E_0
=0$ if and only if $Z\alpha \leq 2/\pi$ (due to Kato [\K] and Herbst
[\H]).  The reason for such a condition is that  the kinetic and
potential energies scale the same way. If the kinetic energy is larger,
$E_0$ can be driven to zero by dilation; in the reverse case, $E_0$ can
be driven to $-\infty$ by contraction. Closely related to this is a
property of the Dirac hydrogenic atom that was known for a long time:
Self adjointness of the Dirac operator requires $Z\alpha \leq 1$.

The stability of the many-body Hamiltonian (for all $N$ and $K$) was
first shown by Conlon [\C] for $Z=1$ and $\alpha < 10^{-200}$ and zero
magnetic field. The situation was substantially improved in [\LiY] to
$Z\alpha \leq 2/\pi$ and $\alpha \leq 1/94$, but $\A=0$.  [\LiY] also
contains a simpler proof of stability which holds only up to a slightly
smaller value of $Z\alpha$ but which has the advantage of holding with
an arbitrary magnetic field. It is this slightly weaker, but more
verstile version of the stability of relativistic matter that enters the
proof of the stability of the Pauli Hamiltonian {\it with} the field
energy $H_{\rm field}$.  (The ability to add a field was not explicitly
stated in [\LiY], but the extension is easy; it was first noted in
[\LLSA].) The constants obtained in [\LiY] have been  improved recently
in [\LLSB]. However, it remains true that the optimal value of $Z\alpha
= 2/\pi$ has not been reached with a magnetic field present. The
sufficient condition for stability in [\LLSB] (with a field) is

$$
({\pi \over 2})Z + 2.80\ Z^{2/3} + 1.30 \leq 1/\alpha \ , \eqno\eqlbl\al
$$
which permits $Z\leq 59$ for $\alpha =1/137$.

We are now ready to use (\al) in the study of the stability of the
nonrelativistic magnetic Hamiltonian (\af). With $Z$ fixed, let us
define the number $\tilde \alpha$ by (\al), i.e.,
$$
1/\tilde\alpha \equiv ({\pi \over 2})Z + 2.80\ Z^{2/3} + 1.30\ .
\eqno\eqlbl\all
$$
Then from (\aj) we conclude the operator inequality
$$
V_{\rm c} \geq  {1\over \tilde \alpha}\sum_{i=1}^N |\p+\A|_i \ .
\eqno\eqlbl\ba
$$
This inequality can be substituted in (\af), and we see that it 
now suffices
to study the spectrum of the one-body operator
$$
h_\A \equiv \T_\A - {\alpha \over \tilde \alpha} |\p+\A|\ .\eqno\eqlbl\bb
$$
In fact, what we require is a lower bound on the the sum of the 
lowest $N$ eigenvalues
of $h_\A$ (according to the Pauli principle). One can show [\LLSA] that 
this sum is bounded below by $const.\times N -  H_{\rm field}$, as required
for stability, {\it provided that} the conditions given in (\ai), or the
ones mentioned after (\ai), are satisfied.

An inequality that plays a key role in this proof is the
CLR bound on the number of negative eigenvalues for a
single particle in a potential (in $n\geq 3$-dimensions), i.e., for the
Hamiltonian 
$$ 
\tilde h \equiv p^2 + U(x)   \eqno\eqlbl\bc 
$$ 
with $p^2
= -\nabla^2$ as usual and with $U$ some arbitrary potential.  We can
always write $U$ in terms  of its positive and negative parts,
$U(x)=U_+(x) - U_-(x)$, with $U_+(x) \equiv (1/2)\{|U(x)| +
|U(x)\}$ and $U_-(x) \equiv (1/2)\{|U(x)| -
U(x)\}$. Then there is the inequality for the number of 
negative eigenvalues
of $\tilde h$, obtained independently by Cwikel, Lieb and Rosenblum, 
$$
\hbox{\rm Number of negative eigenvalues} \leq L_{0,n}\int U_-(x)^{n/2}
 \ d^nx \eqno\eqlbl\bd
$$
with $L_{0,3}=0.1156$. (The sharp value of $L_{0,3} $ is not known, but
it is not much smaller than this value, obtained by Lieb.)

Closely related to this is a family of similar eigenvalue inequalities,
derived earlier by Lieb and Thirring [\LT], about the power sums of the
negative eigenvalues, $e_j$ of $\tilde h$,
$$
\sum_{e_j<0} |e_j|^{\gamma} \leq L_{\gamma, n} \int U_-(x)^{\gamma
+n/2} \ d^nx \eqno\eqlbl\be
$$
for $n\geq 1$, all $\gamma >0 $ for $n\geq 2$ and all $\gamma \geq 1/2$
for $n=1$. This inequality, for $n=3$ and $\gamma =1$ is important in
the proof [\LT] of the stability of nonrelativistic matter without
magnetic fields.  The case $n=3$ and $\gamma =1/2$ plays a role in the
study of $H_{\rm Dirac}$ below. It is known that $L_{1,3} \leq 0.0404$
and $L_{{1\over 2}, 3}\leq 0.06003$.

\chno=4  \equno=0
\bigskip\bigskip
{\chhdsize IV. Putting it all together: Relativistic matter with 
magnetic fields}
\bigskip

Up to now we have considered the  nonrelativistic Schr\"odinger
equation with a Zeeman term (i.e., with the Pauli operator) and 
showed that the magnetic field energy restores the otherwise lost
stability provided $Z\alpha^2 $ and $\alpha$ itself are small enough
(but well within the range of physical parameters). 
We have also considered the ``relativistic'' Schr\"odinger operator,
with no Zeeman term, and found that stability requires that 
$Z\alpha$ and $\alpha$ be small. The combination of the two might 
be expected to lead to insurmountable difficulties, and it was with some
surprise that we found that these difficulties  can be overcome if 
things are defined properly.

We shall try to combine the two by using a Dirac operator for the
electron kinetic energy. Thus, we define the Dirac operator for a
particle of mass $m$ in a magnetic field by
$$
\D_\A\equiv \pslash +\Aslash +\beta m .\eqno\eqlbl\ca
$$ 
As usual,  $\Oslash \equiv  \alpha \cdot O$, where
$\alpha_i $ and $\beta$ are Dirac matrices.
Our many-body Hamiltonian 
is formally similar to the one in (\af), namely,
$$
H_{\rm Dirac} \equiv \sum_{i=1}^N \D_\A(i) +\alpha V_{\rm c} + 
H_{\rm field}
\ ,.\eqno\eqlbl\cb
$$

There is, however, one very important difference from (\af):
$H_{\rm Dirac}$ is unbounded below because the Dirac operator
itself is unbounded below (recall that the Pauli operator, $\T_\A$ is
not only bounded below, it is positive).  According to Dirac the
remedy is to fill all the negative energy states of the Dirac
operator which (because of the Pauli principle) is the same thing
as saying that the electron wave function must lie in the positive
spectral subspace of the Dirac operator.  

There's the rub! Which Dirac operator are we talking about? There are
at least two significant ones in our problem. One is the free Dirac
operator $\D_0= \pslash +\beta m$. The other is the Dirac operator with
the magnetic field $\D_\A$.  It would seem that the latter is the more
important one since an electron can never get rid of its own magnetic
field. Moreover, the choice $\D_0$,  is {\it not gauge invariant}, 
i.e., the
multiplication of an electron wave function by a spatially varying
phase usually takes a positive energy function into a mixture of
positive and negative energy functions.  The second choice is
manifestly gauge invariant. (To avoid possible confusion it is to be
noted that the Hamiltonian is always given by (\cb); the only question
is what condition to impose on the electron wave function.)

This issue is usually not clearly stated in field theory textbooks,
but whatever one might think about the appropriateness of either
definition the interesting fact is that ``stability'' of matter can
settle the argument.  The following two things are proved in [\LSS].
They  show that one choice is valid and the other is not. (Recall that
$\tilde \alpha$ was defined in (\all).)
\bigskip

\vbox{
\noindent
{\bf THEOREM:} \smallskip
{\it \item{(i).} If the electron wave functions are
restricted to lie in the positive spectral subspace of the free Dirac
operator then the Hamiltonian $H_{\rm Dirac}$ is unstable. More
precisely, for any given values of $\alpha >0 $ and $Z>0$, there are
(sufficiently large) particle numbers $N$ and $K$ for which the bottom
of the spectrum of $H_{\rm Dirac}$ is $-\infty$.  
\medskip

\item{(ii).} If  the electron wave functions are restricted to lie in 
the positive spectral subspace of the Dirac operator $D_\A$ then
$H_{\rm Dirac}$ is stable (i.e., $H_{\rm Dirac} >0$)
provided $\alpha $ and $Z$  are small enough. In particular, if we
define $\alpha_c$ to be the unique solution to the equation
$$
1-(\alpha_c/\tilde \alpha)^2 = (8\pi L_{{1\over 2}, 3} \alpha_c)^{2/3}
\eqno\eqlbl\cc
$$
\item{} then $\alpha \leq \alpha_c$ suffices for stability.}
}

As a final remark, we draw attention to a recent result [\EPS].  If we
consider the Dirac hydrogenic atom {\it without} magnetic field, with
Hamiltonian ${\cal D}_0 -Z\alpha /r$, and if the electron's wave
function is required to lie in the positive spectral subspace of $\D_0$,
the critical value of $Z\alpha$, below which this Hamiltonian is bounded
below,  is {\it raised} from the value $2/\pi$ (which is the
corresponding value for the ``relativistic" Hamiltonian in Sect. 3) to
the value $(Z\alpha)_{\rm crit} = 2/(\pi/2+2/\pi)$.  As yet it  is
unknown to what extent a magnetic field will change this value further.

\biblio
\end